%% file: Paper_v9.tex
\documentclass[10pt,twocolumn]{extarticle}
\usepackage[top=1in, bottom=1in, left=0.75in, right=0.75in]{geometry}
\usepackage{authblk}
\usepackage{booktabs}
\usepackage[usenames,dvipsnames]{color}
\usepackage{graphicx}
\usepackage{amsmath}
\usepackage{subfigure}
\usepackage{bm}
\usepackage{amsfonts}
\usepackage[font={small}]{caption}
\usepackage{bm}
\usepackage{multirow}
\usepackage{epstopdf}
\usepackage{soul}
\usepackage[colorlinks]{hyperref}
\hypersetup{
    urlcolor=RoyalBlue,
    linkcolor=Blue,
    citecolor=BrickRed,
    pdfborder={0 0 0}
}
\input{defs.tex}

\begin{document}

\title{\textbf{Interplay of phase boundary anisotropy and electro-autocatalytic surface reactions on the lithium intercalation dynamics in Li$_{\bfX}$FePO$_{\bm{4}}$ platelet-like nanoparticles}}
\author[1]{Neel Nadkarni\footnote{equal contributions}}
\author[1]{Elisha Rejovitzky$^{*}$}
\author[1]{Dimitrios Fraggedakis} 
\author[2]{Claudio V. Di Leo} 
\author[1]{Raymond B. Smith}
\author[3]{Peng Bai}
\author[1,4]{Martin Z. Bazant\footnote{bazant@mit.edu}}
\affil[1]{Department of Chemical Engineering, Massachusetts Institute of Technology, Cambridge, MA 02139, USA} 
\affil[2]{School of Aerospace Engineering, Georgia Institute of Technology, Atlanta, GA 30332, USA} 
\affil[3]{Department of Chemical Engineering, Washington University at St. Louis, MO 63130, USA} 
\affil[4]{Department of Mathematics, Massachusetts Institute of Technology, Cambridge, MA 02139, USA} 
\date{}
\maketitle

\begin{abstract}
Experiments on single crystal Li$_X$FePO$_4$ (LFP) nanoparticles indicate rich
nonequilibrium phase behavior, such as suppression of phase separation at high
lithiation rates, striped patterns of coherent phase boundaries, nucleation by binarysolid
surface wetting and intercalation waves.
These observations have been successfully predicted (prior to the experiments) by 1D
depth-averaged phase-field models, which neglect any subsurface phase separation.
In this paper, using an electro-chemo-mechanical phase-field model, we investigate
the coherent non-equilibrium subsurface phase morphologies that develop in the $ab$-
plane of platelet-like single-crystal platelet-like Li$_X$FePO$_4$ nanoparticles.
Finite element simulations are performed for 2D plane-stress conditions in the $ab$-
plane, and validated by 3D simulations, showing similar results.
We show that the anisotropy of the interfacial tension tensor, coupled with electroautocatalytic
surface intercalation reactions, plays a crucial role in determining the
subsurface phase morphology. With isotropic interfacial tension, subsurface
phase separation is observed, independent of the reaction kinetics, but for strong
anisotropy, phase separation is controlled by surface reactions, as assumed in 1D
models.
Moreover, the driven intercalation reaction suppresses phase separation during
lithiation, while enhancing it during delithiation, by electro-autocatalysis, in quantitative
agreement with {\it in operando} imaging experiments in single-crystalline nanoparticles,
given measured reaction rate constants.
\end{abstract}

\section{Introduction}
In recent years Li$_X$FePO$_4$ has emerged as a significant cathode material for lithium-ion batteries due to its inherent safety and high-rate capabilities~\cite{kang2009battery,zaghib2012enhanced}. 
Its use in diverse applications, such as portable electronics, electric vehicles, renewable energy storage, drives the need to better understand and modify the material for improved rate capability and longer lifetime. 
The electrochemical performance of olivine Li$_X$FePO$_4$ crystals is limited by the complex phase-separation which dominates lithium intercalation in the material~\cite{malik2013critical}. 
In addition, the large stresses that accompany phase-separation lead to mechanical damage and degradation of the battery~\cite{chen2006electron}. 
Controlling the phase-morphology, which is determined by the state of charge (SOC) and rate of charge history, may provide a way to improve battery performance and lifetime~\cite{chiang2010building}. 
Hence it is important to explore and understand the evolution of the phase-morphology throughout the lithiation process.


Rapid progress has been made in directly visualizing the lithiation process at the nanoscale, namely, \emph{in situ} imaging using correlated x-ray adsorption spectroscopy and high-resolution transmission electron microscopy (HRTEM). Both have revealed the phase behavior of individual nanoparticles in porous electrodes~\cite{chueh2013intercalation} at different cycling rates~\cite{li2014current}. 
Similar methods have been applied to map the phase patterns of platelet shaped nanoparticles aligned in the $ac$ plane~\cite{li2015dichotomy}, 
which indicate the presence of a 5 $n$m thick interface that separates the high and low concentration phases~\cite{nakamura2014phase,ramana2009study}, although, variations in the $b$-axis direction cannot be detected. 
The same concerns also apply to recently developed \emph{in operando} techniques, like x-ray imaging~\cite{lim2016origin} and HRTEM~\cite{Zhang2018}, for visualizing the evolution of Li intercalation in platelet-like nanoparticles and provide the most detailed picture of non-equilibrium phase behavior in the $ac$ plane to date.

Few experiments have imaged the concentration evolution in the $ab$ plane, although, producing conflicting results. 
Ohmer et. al. conducted scanning transmission x-ray microscopy (STXM) experiments on single crystalline micron-sized particles to reveal a phase boundary thickness ($\sim$100 $n$m) to be an order of magnitude higher than that observed in the $ac$ plane~\cite{ohmer2015phase}. 
In contrast, the HR-TEM experiment performed by Zhu et. al. in micron-sized particles indicates the formation of a sharp interface interspersed with misfitting dislocations during (de)lithiation~\cite{zhu2013situ}. 
The large phase boundary width in the depth hints towards anisotropy in the interfacial thickness, which introduces a second length scale in the bulk. 

Modeling and simulation provide a crucial complement to experimental measurements, and in the case of Li$_X$FePO$_4$, a number of theoretical predictions about nanoscale phase behavior have preceded and motivated experiments~\cite{bazant2013theory}. 
The earliest models of the intercalation process assumed reduced dimensionality of the primary nanoparticles in order to capture the essential physics with fewer parameters and faster computations~\cite{bai2011suppression,singh2008intercalation,kao2010overpotential}, as required for models of porous electrodes with large numbers of interacting particles~\cite{dargaville2013comparison,ferguson2012nonequilibrium,ferguson2014phase,orvananos2015effect}. 
In particular, many experimentally observed features in non-equilibrium phase separation have been successfully predicted by depth-averaged Allen-Cahn Reaction phase-field models~\cite{bazant2013theory} where concentration variations are confined to the $ac$ plane. 
These theoretical predictions include intercalation waves~\cite{bai2011suppression,burch2008phase,singh2008intercalation} (phase boundaries sweeping across the $ac$ surface, as a domino cascade of filling $b$-axis channels~\cite{delmas2008lithium}), size-dependent and temperature-dependent miscibility gaps~\cite{cogswell2012coherency,burch2009size}, striped patterns of coherent and semi-coherent phase separation~\cite{cogswell2012coherency}, nucleation by binary-solid surface wetting~\cite{cogswell2013theory}, and the suppression of phase separation at high discharge (insertion) rates~\cite{bai2011suppression,cogswell2012coherency}.

Suppression of phase separation at high discharge rates is a surprising prediction of phase-field reaction kinetics~\cite{bazant2013theory}, resulting from composition-dependent reaction resistance (or surface overpotential) for ion insertion. 
In recent years, growing evidence for this phenomenon has been provided by \emph{in situ} experiments involving both surface imaging~\cite{niu2014situ} and volume-averaged bulk imaging~\cite{li2014current,laffont2006study,zhang2014rate}, which reveal solid-solution-like behavior above a critical current density, during discharge. 
In contrast, \emph{in operando} experiments have revealed enhancement of phase separation on reversing the current (or delithiation) at large rates in single crystalline LFP nanoparticles~\cite{lim2016origin}. 
Bazant~\cite{bazant2017thermodynamic} recognized the importance of the functional form of the exchange current. 
The auto-inhibitory nature of the reaction kinetics during lithiation and the auto-catalytic effects upon delithiation, were found to be critical in explaining the asymmetric phase behavior observed during charging (de-lithiation) and discharging (lithiation)~\cite{bai2011suppression}. 
This theoretically predicted role of reaction resistance has been confirmed by direct measurements of the exchange current density versus composition along the $ac$ crystal facet by operando imaging of platelet-shaped Li$_X$FePO$_4$ nanoparticles~\cite{lim2016origin}. 

It is important to note that the theory of suppressed phase separation at the surface also does not rule out the possibility of phase separation below the surface, where the surface overpotential cannot directly influence thermodynamic driving forces. 
Indeed, previous phase-field models of bulk phase separation with surface reactions, assuming 1D spherical symmetry~\cite{zeng2013cahn,tang2009model,tang2010electrochemically,kao2010overpotential}, 2D spheroidal symmetry~\cite{di2014cahn} or 2D planar symmetry (with concentration variations confined to the $ab$ plane)~\cite{dargaville2013persistence}, have predicted subsurface phase separation, albeit without including crystal anisotropy, elastic coherency strain, surface wetting and/or thermodynamically consistent reaction kinetics~\cite{bazant2013theory}.
Some three-dimensional simulations have also been performed, although the computational expense limits the particle size, time-dependence and model complexity that can be considered~\cite{anand2012cahn,di2014cahn,chester2015finite}. 

Tang, et. al.~\cite{tang2011anisotropic} performed the first 3D simulations of lithium intercalation in Li$_X$FePO$_4$ under potentiostatic (constant-voltage) conditions. While their model includes the effects of coherency strain, they did not account for surface-wetting phenomena. 
Additionally, potentiostatic conditions are known to not be able to predict some non-equilibrium phase morphologies within the spinodal gap of voltage vs. SOC curve of a nanoparticle~\cite{bai2011suppression,cogswell2012coherency,cogswell2013theory}. 
Recent 3D modeling efforts have been directed towards predicting equilibrium morphologies in nanoparticles~\cite{welland2015miscibility}, but the findings may not be relevant for the high rates and larger particle sizes that arise in practical applications. 
By studying the lithium equilibrium configurations in FePO$_4$ via \emph{ab initio} techniques coupled with continuum elasticity theory, Abdellahi et. al.~\cite{AbdellahiEtAl2014} were able to explain experimental observations of solid-solution particles~\cite{chen2007metastable}, but they did not account for surface wetting, non-equilibrium lithiation kinetics and anisotropy of the interfacial tension tensor. 
In addition to continuum modeling, Kinetic Monte Carlo (KMC)~\cite{Xiao2018}, have been developed to study the charge and discharge processes, at short timescales, in the $ab$-plane. 
In particular, Xiao and Henkelman~\cite{Xiao2018} implemented KMC for studying the (de)lithiation of FePO$_4$ particles under applied voltage. 
The model captured the essential details at the atomistic scale, however, only described the ideal crystal diffusion resulting in extremely fast (dis)charge rates corresponding to effective C-rates of the order $O\left(10^6\right)$.

In this article, we analyze the morphological changes that develop in Li$_X$FePO$_4$ nanoparticles during constant current lithiation through finite element simulations of a two-dimensional Cahn-Hilliard Reaction model in the $ab$ plane. 
Following earlier work~\cite{bazant2013theory,cogswell2012coherency,di2014cahn}, we develop an electro-chemo-mechanical theory for the intercalation of lithium within FePO$_4$, which is a highly anisotropic phase-separating material~\cite{maxisch2006elastic}. 
The fully coupled theory accounts for the surface wetting properties of the particles along with the deformation and stress generation due to the lithium content, and their effect on both lithium intercalation and diffusion and surface wetting effects, under galvanostatic (constant-current) conditions. 
Additionally, as recent experimental observations indicate larger phase boundary thicknesses in the depth~\cite{zhu2013situ}, we study the effects of an anisotropic phase boundary thickness on the different non-equilibrium phase morphologies during lithiation (i.e. during battery discharge) of Li$_X$FePO$_4$ nanoparticles of a representative thickness of 30 $n$m. 
In particular, we investigate how the anisotropy and the surface dewetting interact with each other to affect the reaction kinetics at the surface which in turn governs the subsurface phase morphology.
A study the effect of imperfections in the single crystalline nanoparticle is left for future work.

\section{Theory}

\subsection{Continuum Model}
We employ a coupled chemo-mechanical Cahn-Hilliard-type continuum model~\cite{anand2012cahn,di2014cahn} with linear elasticity to predict the intercalation in a Li$_X$FePO4 nanoparticle with thermodynamically consistent electrochemical reaction kinetics~\cite{bazant2013theory}. The state of each material point in the nanoparticle is defined by the concentration field $c$, such that $0 < c < c_\text{max}$, where $c_\text{max}$ denotes the maximum theoretical capacity of intercalated lithium in FePO$_4$ matrix, and the strain tensor field $\bm\varepsilon$. We decompose the strain into an elastic strain $\bm\varepsilon^e$ and a chemical strain $\bm\varepsilon^c$, thereby providing the relation $\bm\varepsilon=\bm\varepsilon^e+\bm\varepsilon^c$, where Vegard's law~\cite{Cahn1961} is applied to account for the chemical strains:
\be
	\bm\varepsilon^c = \bm\varepsilon^0 \bar{c}, 
\ee
where $\bar{c} = c/c_\text{max}$ is the normalized concentration, or alternatively the site filling fraction. 
The form of the Helmholtz free energy $F$ of the material in a nanoparticle of domain $B$ 
is decomposed into two parts, a bulk free energy ($F_\text{bulk}$) and a surface free energy ($F_\text{surf}$),
\be
\label{eq:free_energy}
	F = \underbrace{\int_B (f^c(\bar{c})+ f^e(\bm\varepsilon^e)+f^{CH}(\nabla \bar{c})) dV}_{F_\text{bulk}} + \underbrace{\int_{\partial B} \gamma(\bar{c}) dA}_{F_\text{surf}},
\ee
where $f^c$ is the homogeneous chemical free energy density, $f^e$ is the elastic energy density, $f^{CH}$ is the Cahn-Hilliard phase-boundary energy density, and $\gamma$ is the surface energy density. 
For the chemical energy density $f^c$, we employ a regular solution model~\cite{huggins2008advanced}, which has been found to capture quantitatively the bulk thermodynamics of Li$_X$FePO$_4$ ~\cite{bai2011suppression,cogswell2012coherency},
\be
	f^c/c_{\text{max}} = RT (\bar{c}\ln\bar{c} + (1-\bar{c})\ln(1-\bar{c})) + \Omega\, \bar{c}(1-\bar{c}),
\ee 
where $\Omega=4.51RT_m$ is the regular solution parameter taken from the literature~\cite{bai2011suppression,cogswell2012coherency}, $R$ is the universal gas constant, and $T_m = 298 \text{K}$. The elastic energy is taken in the classical form for linear elasticity~\cite{khachaturyan2013theory},
\be
	f^e = \tfrac{1}{2}\bm\varepsilon^e : \left(\mathbb{C} :\bm\varepsilon^e\right),
\ee  
where $\mathbb{C}$ is the fourth-order elasticity tensor which generally depends on the concentration, \cite{AbdellahiEtAl2014}. Herein, this effect is neglected, and the concentration-averaged values, taken from first principle calculations, are used~\cite{maxisch2006elastic}.

The classical Cahn-Hilliard (or Van der Waals) gradient energy~\cite{cahn1958free,Rowlinson1979}, which describes the interfacial tension of a diffuse phase boundary, may be written as follows for anisotropic materials, 
\be
	f^{CH} = \tfrac{1}{2} c_{\text{max}} \nabla \bar{c} \cdot \left(\bfkappa \cdot \nabla \bar{c}\right),
\ee
where $\bfkappa$ is a second-order interfacial tension tensor that is related to the directional phase boundary thickness $\lambda(\bfm)$ through the relation
\be\label{eq:pbt}
	\lambda(\bfm) \propto \sqrt{\frac{\bfm\cdot\left(\bfkappa\cdot\bfm\right)}{\Omega}},
\ee 
where $\bfm$ denotes the direction vector normal to the phase boundary plane~\cite{cogswell2012coherency}. 
The $ac$-plane phase boundary thickness as observed in experiments~\cite{suo2012highly,laffont2006study} and successfully implemented in phase field simulations is taken to be approximately 5~$n$m~\cite{bai2011suppression,cogswell2012coherency,tang2011anisotropic,di2014cahn}. 
Recent experimental studies~\cite{ohmer2015phase} revealed the need for the anisotropic modeling of the phase boundary. 
This effect can be incorporated into the modeling part by considering the components of $\bfkappa$ to differ between each other.  
To this end, the phase boundary in the depth direction of the particle ($b$-axis) is considered to be an order of magnitude higher than that corresponding to the $ac$-plane.
We compare our results with that of an isotropic interfacial tension tensor corresponding to a phase boundary of 5~$n$m. 
The exact values for the parameters are provided in the Supplementary Materials.
%

Taking the variational derivatives of the free energy,~\eqref{eq:free_energy}, with respect to the concentration and the strain yields the diffusional chemical potential, $\mu$, and the activity, $a$, of the diffusing lithium
\be
\label{eq:chem_pot}
\begin{split}
	&\mu = \mu^0 + RT \ln a = \frac{\delta F}{\delta c} = \mu^0 + RT \ln\left(\frac{\bar{c}}{1-\bar{c}}\right) \\&+ \Omega\, (1-2\bar{c}) - \nabla\cdot(\bfkappa\cdot\nabla\bar{c}) - c_{\text{max}}^{-1} \bfsigma:\bm\varepsilon^0,
\end{split}
\ee
and the elastic stress
\be
\label{eq:stress}
	\bfsigma = \frac{\delta F}{\delta \bm\varepsilon} = \mathbb{C}: (\bm\varepsilon-\bm\varepsilon^0\bar{c}), 
\ee
respectively. Note that the coupling between the chemical potential and the stress~\cite{christensen2006stress,cogswell2012coherency} is expressed by the term $-c^{-1}_\text{max}\bfsigma:\bm\varepsilon^0$ in~\eqref{eq:chem_pot}, and by $\bm\varepsilon^0\bar{c}$ in~\eqref{eq:stress}. 

The balance laws for the system, or the governing partial differential equations of the problem are (i) balance of forces, and (ii) balance of species. Neglecting body forces, the balance of forces requires that, 
\be
	\nabla\cdot\bfsigma = 0.
\ee
Mass conservation requires
\be
	\frac{dc}{dt} = -\nabla\cdot \bfj,
\ee
where the flux density~\cite{bazant2013theory,nauman2001nonlinear},
\be
	\bfj = - c(1-\bar{c})\bfM \cdot\nabla \mu
\ee
is expressed in terms of a second-order mobility tensor, $\bfM$, related to the diffusivity tensor, $\bfD$, through the Einstein relation, $\bfM=\bfD/RT$. 
The strongly anisotropic diffusivity of LiFePO$_4$~\cite{Morgan2004} favors diffusion along ion-channels in the direction of the $b$-axis, which is expressed in the model by diffusivity in the $b$ direction which is six orders of magnitude higher than in the $ac$ plane. 

Surface energy has a significant effect on the phase-morphology~\cite{cogswell2012coherency} and the performance~\cite{zeng2013cahn} of Li$_X$FePO4. 
Comparing \emph{ab initio} computations of the surface energies of facets of LiFePO$_4$ and FePO$_4$ (exposed to vacuum)~\cite{wang2007first} and the LiFePO$_4$ / FePO$_4$ phase-boundary energy reveals that the $bc$ and $ab$ side-facets have a much lower energy for Li-rich phases and tend to be fully ``wetted" with a surface layer of intercalated lithium ($\bar{c}\approx1$). 
The wetting of the $bc$ and $ab$ facets drives the heterogeneous nucleation of Li-rich phase at the side facets~\cite{bai2011suppression,cogswell2013theory}, as observed in equilibrium phase-morphologies from \emph{ex-situ} measurements~\cite{laffont2006study}. 
In contrast, the vacuum surface energy of the $ac$ facets is much lower for Li-poor phases, which would imply that these correspondingly tend to fully ``de-wet" ($\bar{c}\approx0$). 
This may seem to contradict the fact that the $ac$ facets are the most electrochemically active sites for lithium insertion, and it does alter the activation overpotential for intercalation reactions. 
However, it is likely that carbon coatings in practical battery particles alter surface activity (as well as electronic conductivity) so as to enhance lithium wetting of the active facets. 
In any case, we will use \emph{ab initio} surface energy versus vacuum, which leads to de-wetting of the active $ac$ facet, in order to maximize the possibility of subsurface phase separation in the depth $b$ direction, by providing surfaces for heterogeneous nucleation.

The functional derivative of the total free energy with respect to concentration yields, in addition to the bulk chemical potential~\eqref{eq:chem_pot}, the natural surface-wetting boundary condition~\cite{bazant2013theory}:
\be
\label{eq:bcWetting}
	\bfn\cdot\left(\bfkappa\cdot\nabla \bar{c}\right) = -\frac{1}{c_{\text{max}}}\frac{\partial \gamma(\bar{c})}{\partial \bar{c}},
\ee
where $\gamma(\bar{c})$ and $\bfn$ are the surface energy and the outward normal vector to the surface respectively. The values of the surface energy of the Li-rich and Li-poor phase ($\gamma_{\text{LiFePO$_4$}}, \gamma_{\text{FePO$_4$}}$) are taken from \emph{ab initio} computations~\cite{wang2007first} and extrapolated with the function,
\be
	\gamma(\bar{c}) = \Delta \gamma(3\bar{c}^2-2\bar{c}^3) + \gamma_{\text{FePO$_4$}},
\ee
where $\Delta\gamma = \gamma_{\text{LiFePO$_4$}}-\gamma_{\text{FePO$_4$}}$. 
All the parameters used for the continuum model are summarized in Table 1 in the Supplementary Materials.

\subsection{Reaction model}
The rate-determining step of the electrochemical reaction is assumed to be
\begin{equation*}
	\text{Li}^+ + \text{e}^- + \text{FePO}_4 \to \text{LiFePO}_4
\end{equation*}
while the exchange of lithium between the electrolyte and the Li$_X$FePO$_4$ occurs only on the $ac$ facets which are open to the ion channels, \cite{Morgan2004}. We use the symmetric case of the Butler-Volmer equation~\cite{bazant2013theory} to describe the reaction and implement it via a surface flux boundary condition,
\be
\label{eq:bcCurrent}
	\bfj\cdot\bfn = 2\, j_0(\bar{c}) \sinh\left(\frac{\mu-F\phi}{RT}\right)
\ee
where $\phi$ and $F$ are the applied voltage and the Faraday constant respectively.  Although the voltage-dependence of the Butler-Volmer equation breaks down at high overpotentials and may need to be replaced by quantum mechanical electron-transfer theory~\cite{bai2014charge}, we focus here instead on the concentration dependence of the exchange current, which more directly influences phase separation under applied current~\cite{bazant2017thermodynamic}, while keeping the standard Butler-Volmer voltage dependence. 

We consider two fundamentally different models for the exchange current density $j_0(\bar{c})$. The first is the standard empirical formula used in battery modeling, proposed by Doyle, et. al.~\cite{doyle1993modeling}:
\be
\label{eq:rm_Doyle}
	j_0(\bar{c}) = k\sqrt{\bar{c}(1-\bar{c})}
\ee
which is symmetric with a peak at $\bar{c}=0.5$. The second is an asymmetric profile obtained  by direct {\it in operando} X-ray imaging experimental measurements by Lim et. al.~\cite{lim2016origin}:
\be
\label{eq:rm_Lim}
	j_0(\bar{c}) = 3k(1-\bar{c})\sqrt{\bar{c}(1-\bar{c})}
\ee
which is shifted to lower concentrations and peaks around $\bar{c}\sim0.2$. The concentration dependence of this formula is similar to that predicted earlier by Bazant~\cite{bazant2013theory} for generalized (symmetric) Butler-Volmer kinetics with one excluded site in the  transition state,
\be
\label{eq:rm_Bai}
	j_0(\bar{c}) = k\sqrt{a} (1-\bar{c})
\ee
which has been used successfully to predict driven phase separation in depth-averaged phase-field models of LFP ~\cite{bai2011suppression,cogswell2012coherency,cogswell2013theory}, including effects of coherency strain.   Recently, it has been discovered that the experimentally measured concentration dependence of the exchange current, Eq. (\ref{eq:rm_Lim}) can be predicted (without any fitting parameters) by an analogous generalization of Marcus-Hush-Chidsey (MHC) kinetics for Faradaic reactions at electrodes~\cite{smith2018electron}, which combines the quantum mechanical theory of electron transfer~\cite{marcus1956theory,marcus1965electron,dogonadze1965theory,chidsey1991free,zeng2015simple} with the nonequilibrium thermodynamics of ion intercalation~\cite{bazant2013theory}.  

The theory of driven surface phase separation under an applied current~\cite{bazant2017thermodynamic}, previously applied to depth averaged models, predicts that phase separation is suppressed during insertion, if the reaction is auto-inhibitory across the spinodal region of intermediate concentrations. This requires an exchange current $j_0(\bar{c})$ that is asymmetric around $\bar{c}=0.5$ and peaked at low $\bar{c}$ values, in which case, the theory also predicts that phase separation is enhanced by electro-autocatalysis when the current is reversed during extraction.  (See Fig. 3 in Ref.~\cite{bazant2017thermodynamic}.) These predictions have been directly verified by visualizing the phase behavior of individual LFP nanoparticles during insertion/extraction cycles at different rates, coupled with local, nanoscale measurements of the reaction-rate far from equilibrium~\cite{lim2016origin}.  The rate-dependent control of phase separation was shown to be consistent with the asymmetric profile of   $j_0(\bar{c})$  in Eq. (\ref{eq:rm_Lim}) from the experiments and theoretical models~\cite{bai2011suppression,smith2018electron}, but not with the commonly assumed symmetric profile, Eq. (\ref{eq:rm_Doyle}), thus providing a compelling test of the theory. Moreover, direct observations of surface nucleation and striped phase patterns in single nanoparticles~\cite{lim2016origin}, consistent with predictions by the same theory~\cite{cogswell2012coherency,cogswell2013theory}, dispel earlier claims of a "solid-solution pathway" in LFP resulting from suppressed nucleation~\cite{kang2009ultrafast}, and establish the crucial role of coherency strain in nanoparticle phase separation.  Here, we study how these nonequilibrium surface phenomena are affected phase separation dynamics below the surface. 

The influence of driven interfacial reactions on phase separation depends on the magnitude of the rate constant relative to diffusion times (Damk\"{o}hler numbers), and the imposed current~\cite{singh2008intercalation,bazant2017thermodynamic}. Estimated rate constants for lithium insertion in LFP greatly vary in the literature (from $< 10^{-5}$ A/m$^2$ \cite{srinivasan2004discharge} to  $>10$ A/m$^2$ \cite{wang2007discharge,kasavajjula2008discharge}). Although there are differences in electrode preparation methods, the huge discrepancy is more likely attributable to the choice of mathematical model used to interpret the data and to the unknown active surface area of a porous electrode undergoing phase separation.  Classical porous electrode theory, which assumes radial solid diffusion in active particles and fits the open circuit voltage rather than the non-convex free energy surface, greatly over-estimates the active area at low rates and thus infers tiny rate constants, such as $k=3 \times 10^{-6}$ A/m$^2$~\cite{srinivasan2004discharge}  and 5.4$\times 10^{-5}$ A/m$^2$ ~\cite{dargaville2010predicting}.  In contrast, multiphase porous electrode theory (MPET)~\cite{ferguson2012nonequilibrium,smith2017multiphase}, which accounts for surface intercalation waves~\cite{singh2008intercalation,tang2011anisotropic,bai2011suppression,cogswell2012coherency} and rate-dependent, reduced populations of active particles (``mosaic instabilities")~\cite{dreyer2010thermodynamic,chueh2013intercalation,li2015dichotomy,bai2013statistical,li2014current}, infers much larger rate constants for LFP intercalation, such as $k=7\times 10^{-3}$ A/m$^2$ ~\cite{ferguson2014phase}.   This value is quantitatively consistent with the best available measurement, $k\sim 10^{-2}$ A/m$^2$, from {\it operando} x-ray imaging of individual nanoparticles by Lim et al.~\cite{lim2016origin}. These experiments also revealed fast and slow domains at the nanoscale, where the rate constant may differ within an order of magnitude, presumably due to surface heterogeneities.  Somewhat slower local reaction rates ($k \sim 5\times10^{-4}$ A/m$^2$) have been measured experimentally by micro-diffraction experiments~\cite{zhang2015direct}, consistent with the estimate of Bai and Bazant~\cite{bai2014charge} by fitting chronoamperometry data to a simple population dynamics model with MHC reaction kinetics ($k \sim 10^{-4}$ A/m$^2$).  In light of all of these experimental and theoretical results, we will assume a realistic rate constant for ``slow domains" in the experiments of Lim et al.~\cite{lim2016origin}, $k = 10^{-3}$ A/m$^2$, which allows us to accurately predict the observed C rates for suppression of phase separation, although the qualitative results of our study are insensitive to the precise value of the rate constant. 

In order to model capture nonequilibrium morphologies, we simulate galvanostatic discharging by controlling the total insertion flux of lithium into the nanoparticle by adjusting the applied voltage~\cite{bai2011suppression,cogswell2012coherency,cogswell2013theory}. Constant total current is implemented through the following integral boundary constraint,
\be
\label{eq:cst_I}
	I = F\int_{\partial B_{\text{reac}}}\bfj\cdot\bfn \,dA.
\ee
where $\partial B_{\text{reac}}$ represents the active top and bottom $ac$ reaction facets. Equation \eqref{eq:cst_I} serves as the constraint which determines the resulting voltage under constant current $I$. 

\subsection{Particle geometry and boundary conditions}
\begin{figure}[t]
\centering
\includegraphics[width=2.7in]{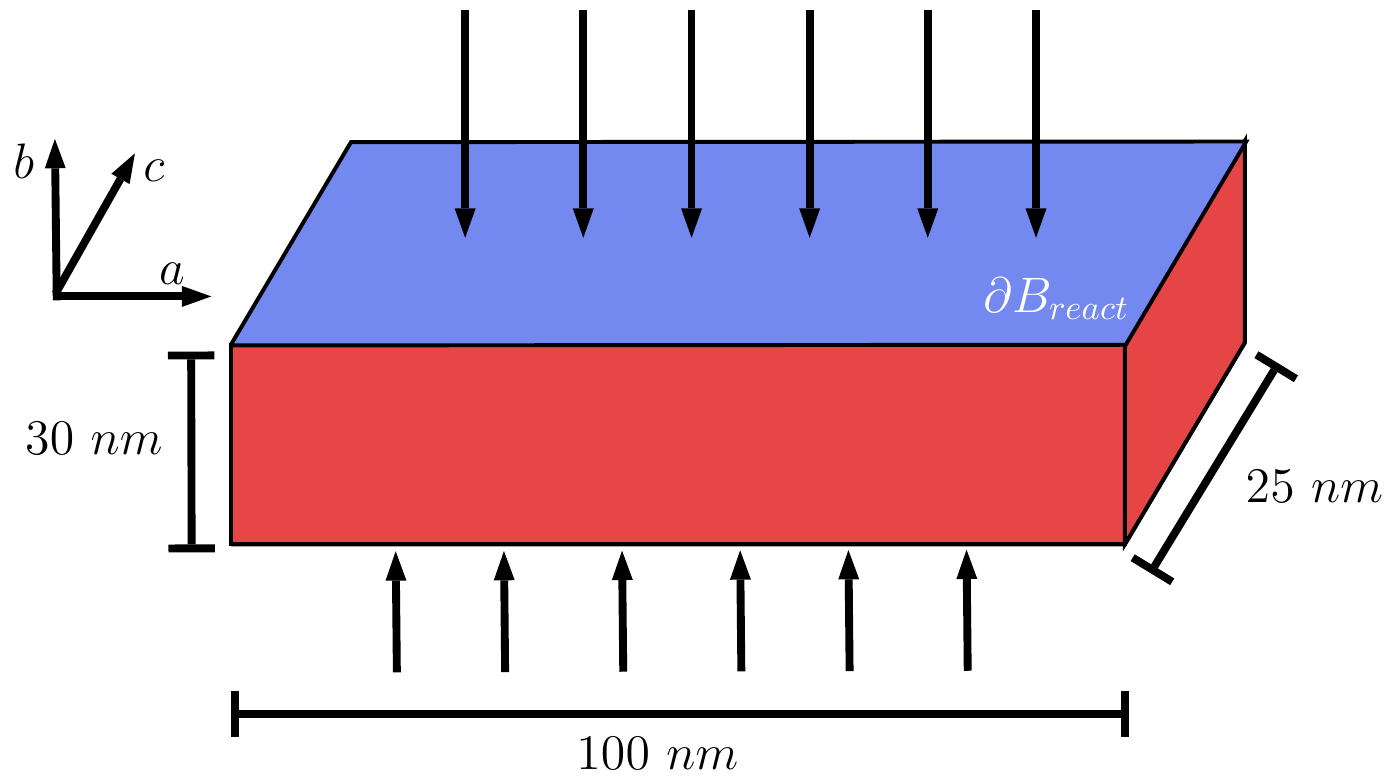}
\caption{Schematic of a FePO$_4$ particle. $Li^{+}$ intercalates through $\partial B_{react}$ sides.}
\label{fig:cad}
\end{figure}
In the following section we analyze the phase-morphology in platelet-like nanoparticles. 
Fig.~\ref{fig:cad} illustrates the representative 3D shape of a FePO$_4$ particle studied in the present work. In order to reduce the computational effort while retaining the main morphological characteristics, we simplified the nanoparticle geometry to 2D, where the platelet shape is represented by a rectangular domain of the $ab$ plane. 
The main reason for such simplification lies on the fact that we are interested on the conditions under which phase separation is prevented in the $b$ direction.
As the thickness of the particle in the $c$ axis is relatively small as compared to the $a$ axis, a 2D plane stress approximation is used. 
The assumption of the plane-stress case is further validated by simulating the full 3D particle, under similar lithiation conditions. 
and we show results for one particular case. 

The nanoparticles were chosen to have a length of 100 $n$m (in the $a$-axis), thicknesses of 30 $n$m (in the $b$-axis), and for the 3D case a representative length of 25 $nm$ in the $c$-direction was considered. 
The galvanostatic integral constraint was imposed by integrating the flux over the combined top and bottom ac facets, eq.~\eqref{eq:bcCurrent}, and zero-flux was applied on the sides. 
Wetting and dewetting boundary conditions, eq.~\eqref{eq:bcWetting}, were applied on the $bc$ and $ac$ facets respectively. 
In practice, the LiFePO$_4$ nanoparticles are mechanically constrained by their carbon coating, the binder, and contact with the current collector and other nanoparticles. 
Such constraints may affect the lithiation process, work to suppress phase separation, and, along with mechanical defects in the nanoparticle, may determine nucleation sites. 
In this study we neglect the mechanical interactions and prescribe zero traction boundary conditions on all facets of the nanoparticle, i.e. $\bfsigma\cdot\bfn = \bfnull$ on the surface. The rigid modes of rotation and translation are eliminated by pinning the bottom left corner and prescribing a zero $b$-axis displacement to the bottom right corner. 

\subsection{Characteristic length and time-scales}
Following Singh et. al.~\cite{singh2008intercalation}, the diffusivity, exchange current, and geometry of the nanoparticle define four time-scales which help in characterizing the evolution of the phase-morphology, and are related to different physical mechanisms: (i) the lithiation time $\tau_C$ is the time it takes the nanoparticle to fully lithiate at a given
current; (ii) the characteristic diffusion time inside the ion channels $\tau_D = h^2/D_b$ represents the time it takes the species to diffuse across the ion channels, where $h$ is the nanoparticle thickness; (iii) the characteristic reaction time $\tau_R = c_{\text{max}}Fh/k$ describes the time it takes the reaction to fill the ion channels; and (iv) the
characteristic diffusion time in the $a$ direction $\tau_E = l^2/D_a$ defines the time for lithium exchange across a $bc$-oriented interface, where $l$ is the interface width.
The four characteristic time scales are further reduced to a set of three non-dimensional parameters.
The Damk\"{o}hler number in the $b$ direction is defined as $Da_b = \tau_D/\tau_R \sim 10^{-7} \ll 1$ and shows that the lithiation process is reaction limited, as assumed in depth-averaged models~\cite{bai2011suppression,cogswell2012coherency,singh2008intercalation}. The $a$-axis non-dimensional Damk\"{o}hler number $Da_a= \tau_E/\tau_R \sim 3\cdot10^{-4}$ shows that the $a$-axis diffusion which is typically neglected~\cite{bai2011suppression,cogswell2012coherency,singh2008intercalation,tang2011anisotropic}, is in fact non-negligible. 
Finally, the non-dimensional current $I = \tau_R/\tau_C$ is scaled to the inverse reaction time, i.e. the time required to fill the particles at the characteristic reaction rate. 

There are two important interfacial length scales observed in this system: (i) the length scale 
\be
\label{eq:int_thck}
	t(\bfm) = 3.21\sqrt{\frac{\bfm\cdot\left(\bfkappa\cdot\bfm\right)}{\Omega}},
\ee 
associated with the thickness of the phase boundary oriented in a direction perpendicular to unit vector $\bfm$, where 3.21 is the constant of proportionality in~\eqref{eq:pbt}. In the case of isotropic $\bfkappa$, $t(\bfm)$ is independent of the interface orientation, and (ii) the length scale 
\be
	l_s = \frac{\kappa_y c_\text{max}}{6\Delta\gamma}, 
\ee
corresponding to the thickness of the interfacial layer formed at the surface due to dewetting along the $ac$-plane. As we shall see, the comparison of $l_s$ with the thickness of the interface in the depth $t(\bfe_b)$ or, in the other words, the ratio $l_s/t(\bfe_b)$ will govern the concentration of the surface layer that is formed due to surface dewetting at the boundary, which in turn will affect the reaction kinetics at the surface. Similarly, the ratios $t(\bfe_a)/L$ and $t(\bfe_b)/h$, where $\bfe_a$ and $\bfe_b$ are the unit vectors in the direction of the $a$ and $b$ axes respectively, will determine if an interface will be formed along the $a$ and $b$ axes. The following section presents results for non-equilibrium phase behavior based on an isotropic and an anisotropic $\bfkappa$. The anisotropy of $\bfkappa$ is such that $t(\bfe_b)$ is chosen to be 10 times that of $t(\bfe_a)$, an order of magnitude higher, as indicated by experiments~\cite{ohmer2015phase}. 
\begin{figure}[t]
\centering
\includegraphics[width=3.6in]{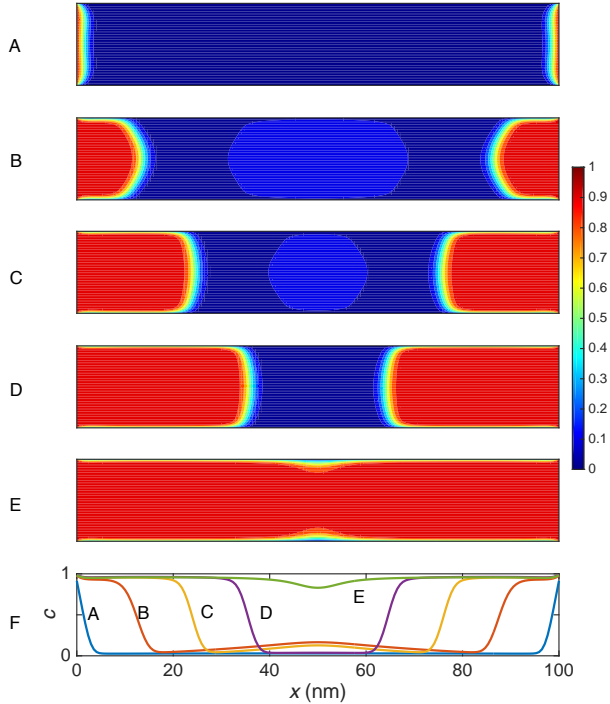}
\caption{(A)-(E) Concentration profiles in the nanoparticle during slow lithiation of 0.1C at SOC's with an isotropic $\bfkappa$ tensor for an average concentration, $X$ of: (A) 0.05, (B) 0.3, (C) 0.5, (D) 0.7, and (E) 0.95. (F) Depth averaged concentration profiles of (A)-(E).}
\label{fig:1}
\end{figure}

\section{Simulations for small and medium currents}
\subsection{Isotropic interfacial thickness}
We first conduct our numerical simulations with an isotropic interfacial tension tensor. 
Different stages during a lithiation process at 0.1C rate for the symmetric reaction model~\eqref{eq:rm_Doyle} are presented in Fig.~\ref{fig:1}, starting with an almost empty nanoparticle with nucleated Li-rich regions at the side facets (Fig.~\ref{fig:1}A). (The C-rate, $x$C, is defined as the current required to fully lithiate the particle in 1/$x$ hours.) 
As lithiation progresses, the lithium atoms are directed towards the regions with the greatest chemical potential difference, which are those near the phase interfaces, thereby creating intercalation waves~\cite{burch2008phase,delmas2008lithium} and thickening the Li-rich regions at the side facets (Fig.~\ref{fig:1}B). 
The dewetting effect at the top and bottom surfaces combined with elastic effects yields a semi-circular shape of the Li-rich regions. 
The effect of mechanics can be understood by considering a straight interface which is normal to the $a$-axis.
In that morphology, the Li-poor region next to the interface is under tension in the $b$-axis which vanishes at the top and bottom surfaces and reaches a maximum in between. 
The tension reduces the chemical potential through the coupling and drives the lithium to balance the chemical potential by curving the interface.
At higher SOC's (Fig.~\ref{fig:1}C) the Li-rich regions further thicken and move towards the center of the nanoparticle with an interface normal to the $a$-axis direction, corresponding to the energetically preferable interface orientation at equilibrium~\cite{cogswell2012coherency}. 
In addition, the strong dewetting properties of the top and bottom facets keep them at a low concentration which maintains a constant exchange current density locally.
As lithiation proceeds (Fig.~\ref{fig:1}D), the two Li-rich regions coalesce, and further insertion of lithium occurs through filling up the regions of low concentration near the top and bottom facets (Fig.~\ref{fig:1}E).
Depth averages of the concentration profiles in Fig.~\ref{fig:1}A-E are presented in Fig.~\ref{fig:1}F and show good agreement with previous results of the depth averaged model~\cite{bai2011suppression,cogswell2012coherency,singh2008intercalation}, in which the assumption of uniform concentration in each ion channel was used. 
The slow lithiation process follows through near-equilibrium morphologies that, far enough from the wetted edges, minimize the elastic energy by developing interfaces normal to the $a$-axis, in agreement with the findings by Cogswell and Bazant~\cite{cogswell2012coherency} for the equilibrium case.
\begin{figure}[t]
\centering
\includegraphics[width=3.6in]{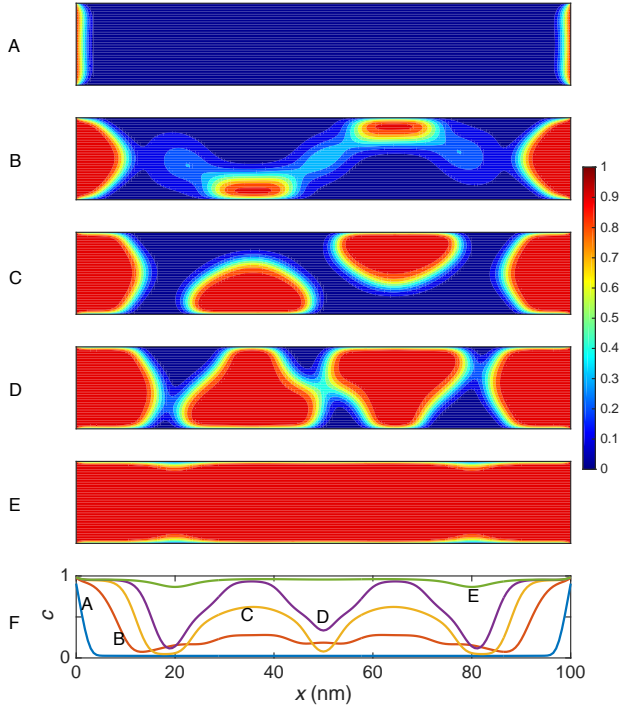}
\caption{(A)-(E) Concentration profiles in the nanoparticle during fast lithiation of 1C at SOC's with an isotropic $\bfkappa$ tensor for an average concentration, $X$ of: (A) 0.05, (B) 0.3, (C) 0.5, (D) 0.7, and (E) 0.95. (F) Depth averaged concentration profiles of (A)-(E).}
\label{fig:2}
\end{figure}

Fast lithiation results in phase-morphologies which are further away from equilibrium than the ones observed during slow lithiation. 
The concentration profile at different stages of a 1C lithiation rate process are presented in Fig.~\ref{fig:2}. 
As the nanoparticle fills up from its initial state (Fig.~\ref{fig:2}A), the nucleated Li-rich regions at the side facets thicken as in the low lithiation rate. 
In addition to that, to sustain the large current, the concentration and chemical potential at the center of the nanoparticle increase (Fig.~\ref{fig:2}B) up to the spinodal point. 
Then, the center of the nanoparticle undergoes spinodal decomposition and separates into a Li-rich and a Li-poor phase (Fig.~\ref{fig:2}C). 
The decomposition occurs rapidly through lithium diffusion along the ion channels, as recently predicted for spinodal decomposition simulations without lithiation/delithiation~\cite{abdellahi2016thermodynamic}. 
Recall that the mechanism of diffusion along the ion channels is much faster than species exchange between the ion channels through either reaction or diffusion in the $a$-axis direction. 
The species diffuse to the top and bottom facets of the nanoparticle and form an island of Li-rich phase which is almost uniform in the $a$-axis direction and has an interface in the $b$-direction. 
High stresses develop at the interface as a result of the high chemical strains in the $a$-axis direction ($\varepsilon^0_{aa}$), and thicken the interface. 
As lithiation progresses, the interfaces formed in the depth move in a direction parallel to the $b$-axis and eventually reach the surface while coalesce in the middle. 
Note that the interface normal to the $b$-axis that appears during high lithiation rates is far from an equilibrium morphology.
Depth averages of the concentration profiles in Fig.~\ref{fig:2}A-E are presented in Fig.~\ref{fig:2}F. 
The moving front normal to the $b$-axis, as seen in Fig.~\ref{fig:2}B-D, presents itself as a spinodal decomposition starting at the spinodal point (B) through the spinodal gap (C) until the particle is full (E). 

An important point to note is that, as the concentrations on the top and bottom facets remain almost fixed during the whole duration of the reaction, the form of the reaction model does not affect the kinetics of phase separation in the bulk. 
Hence, the phase morphologies obtained for both reaction models show little variation. 
The effects of reaction kinetics are seen in the case of an anisotropic interfacial thickness, the results of which are presented in the following section.

\subsection{Anisotropic interfacial thickness}
In this section, we report our results for an anisotropic $\bfkappa$ tensor for different insertion rates. 
We first highlight the distinctions in the thermodynamics of the anisotropic case as compared to that of the isotropic case that are independent of the reaction rate. 
There are two fundamental differences as compared to the isotropic case: (i) the curvature of the phase boundary is smaller than that of isotropic case and the concentration is uniform in the depth irrespective of the kinetics of the reaction, (ii) the concentration along the $ac$ facet is unaffected by the surface dewetting unlike the isotropic case where the surface is almost fully dewetted $(c \approx 0)$.

The first observation can be explained by comparing the expected interfacial thickness in the $b$ direction with the actual thickness of the particle. 
In particular using Eqn.~\eqref{eq:int_thck} it is found that the interface in the $\bfe_b$ direction has approximately $\sim50\, n$m width, which clearly exceeds the height of the particle. 
By taking into account the elastic effects too, the interfacial width is expected to be even larger for the selected $\kappa_y$ value. 
In general, it is well known that phase separation can be eliminated for particles which have dimensions less than interface width~\cite{Cahn1961}, while this effect becomes even more pronounced under non-equilibrium operation conditions~\cite{burch2009size}.

The second observation can be analyzed through the scaling relation for the natural boundary condition for the concentration on the top and bottom surfaces given by,
\be\label{eq:scaling}
	\frac{\partial \bar{c}}{\partial y} = -\frac{1}{c_{max}\kappa_y} \gamma'(\bar{c}) 
\ee
Suppose, the normalized concentration at the surface is $\bar{c}_0$ and below the surface a high concentration phase $(\bar{c} \approx 1)$ has formed through spinodal decomposition. Then, the gradient term in~\eqref{eq:scaling} will scale as,
\be
	\frac{\partial \bar{c}}{\partial y} \sim \frac{\bar{c}-1}{t(\bfe_b)},
\ee
while the right hand side of the equation will be given by,
\be
	-\frac{1}{\kappa_yc_\text{max}} \gamma'(c) = \frac{1}{\kappa_y c_\text{max}} 6\left\lvert\Delta\gamma\right\rvert\bar{c}(\bar{c}-1).
\ee
Therefore, the concentration at the surface can be computed by the scaling equation
\be
\label{eq:scalingLaw}
	 \left(\frac{\kappa_y c_\text{max}}{6\left\lvert\Delta\gamma\right\rvert t(\bfe_b)} - \bar{c}\right)(\bar{c}-1) \sim 0
\ee
There are two ways of satisfying~\eqref{eq:scalingLaw}. If 
\be
	\frac{\kappa_y c_\text{max}}{6\left\lvert\Delta\gamma\right\rvert t(\bfe_b)} < 1 \quad \Rightarrow \quad \bar{c} = \frac{\kappa_y c_\text{max}}{6\left\lvert\Delta\gamma\right\rvert t(\bfe_b)}, 
\ee
then, the 
\be
	\frac{\kappa_y c_\text{max}}{6\left\lvert\Delta\gamma\right\rvert t(\bfe_b)} > 1 \quad \Rightarrow \quad \bar{c} = 1. 
\ee
Hence, for a large $\kappa_y$, the gradient in concentration is very small at the boundaries along the $ac$ facet, when phase transformation has occurred in the bulk.

\begin{figure}[t]
\centering
\includegraphics[width=3.6in]{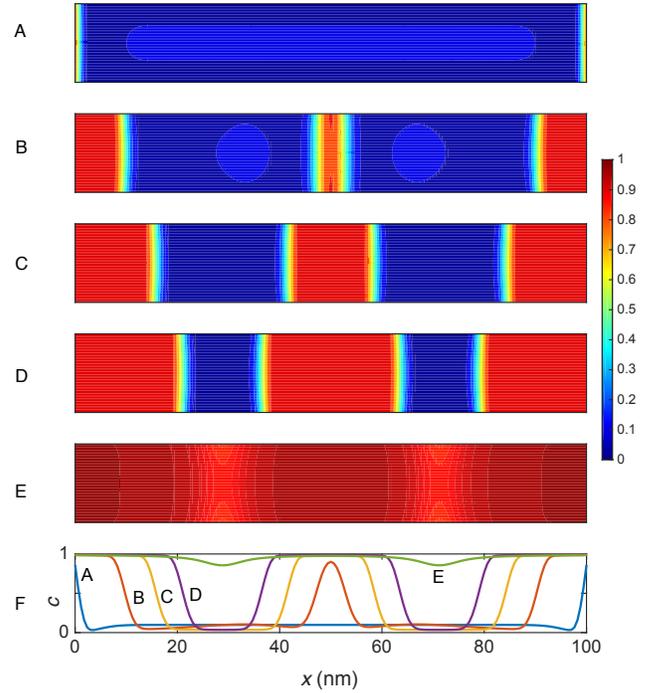}
\caption{(A)-(E) Concentration profiles in the nanoparticle during slow lithiation of 0.1C at SOC's with an anisotropic $\bfkappa$ tensor and reaction model~\eqref{eq:rm_Lim}  for an average concentration, $X$ of: (A) 0.05, (B) 0.3, (C) 0.5, (D) 0.7, and (E) 0.95. (F) Depth averaged concentration profiles of (A)-(E).}
\label{fig:4}
\end{figure}

Next, we look at the effect of reaction rate on the phase morphology observed in the bulk.
Fig.~\ref{fig:4} shows the evolution of the concentration for a slow lithiation rate of 0.1C. 
The concentration here remains uniform across the depth, similar to the isotropic case. 
As the reaction progresses, a phase transformation occurs at the center of the particle through spinodal decomposition and proceeds quickly to become uniform in the depth direction. 
Following the formation of the new phase, the reaction proceeds as a traveling wave similar to that of the isotropic case and as predicted by depth-averaged models~\cite{bai2011suppression,cogswell2012coherency,singh2008intercalation}. 

\begin{figure}[t]
\centering
\includegraphics[width=3.6in]{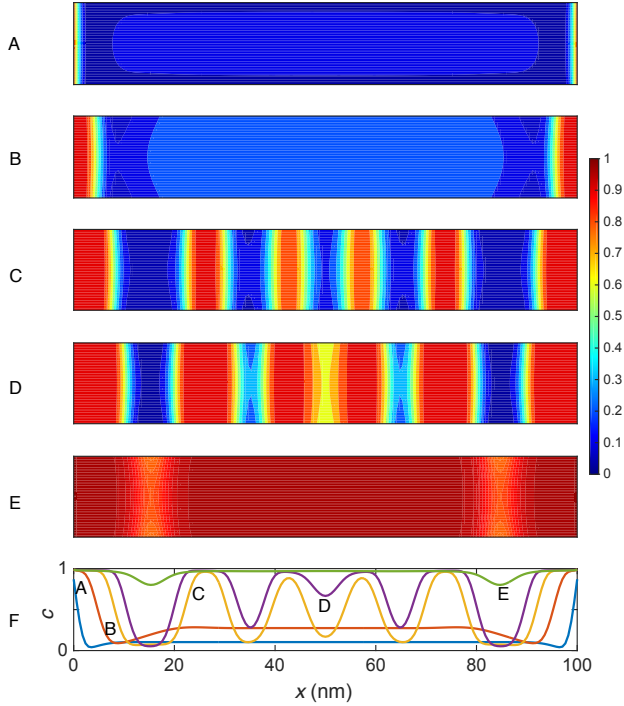}
\caption{(A)-(E) Concentration profiles in the nanoparticle during fast lithiation of 1C at SOC's with an anisotropic $\bfkappa$ tensor and reaction model~\eqref{eq:rm_Lim}for an average concentration, $X$ of: (A) 0.05, (B) 0.3, (C) 0.5, (D) 0.7, and (E) 0.95. (F) Depth averaged concentration profiles of (A)-(E).}
\label{fig:5}
\end{figure}

Fig.~\ref{fig:5} depicts the lithiation process for a large C-rate of 1C. 
As opposed to the isotropic case, as explained earlier, the concentration here remains constant across the depth. 
The phase morphology is very different from the one observed in a case with isotropic interfacial tension and hence the depth averaged concentration also evolves differently as compared to an isotropic case. 
Multiple high concentration regions begin to emerge as a result of spinodal decomposition at different points along the $a$-axis. 
The interface formed, as a result of the phase separation, move in the form of traveling waves coalescing with each other on impact. 
The mechanism is the same as the one corresponding to the slow lithiation process at 0.1C, the difference being that in fast lithiation, a larger number of phase separated regions are created. 
Hence, for an anisotropic interfacial tension tensor, the results can be completely explained by a reduction in dimensionality i.e. using a depth-averaged model. 
The following section analyzes how exactly does the reaction kinetics affect the phase behavior for large C-rates in the isotropic and anisotropic cases.

\section{Simulation results for large currents}
Multiple depth-averaged theories have predicted suppression of phase separation at high lithiation rates in LFP~\cite{bai2011suppression,cogswell2012coherency} and qualitative agreement with recent experiments has been found 
as well~\cite{lim2016origin}. 
An interesting outcome of the depth-averaged theory was the effect of the reaction model at the surface on the dynamics of phase separation. 
In particular, the interplay of diffusion with reaction kinetics, specifically the form of the exchange current density, affects the lithiation of particles significantly, suppressing phase separation at high rates for an auto-inhibitory type of reaction and enhancing it for an auto-catalytic type~\cite{bazant2017thermodynamic}. 
STXM experiments measuring the depth-averaged $ac$-plane concentration in single-crystalline particles confirmed this finding through the observation of asymmetric phase patterns showing an enhanced phase separation during delithiation in comparison to almost homogenous filling during lithiation~\cite{lim2016origin}. 

In order to investigate the exact effect of the surface reactions on the bulk phase morphology, we need to analyze the phase behavior that occurs in the bulk.
A good candidate for quantifying the amount of phase separation in the nanoparticle is the spatial variance of the concentration,
\be
	\langle (c-X)^2 \rangle = \frac{1}{V}\int_V \left|c-X\right|^2 dV,
\ee 
which signifies the variation of the local concentration with respect to the mean concentration, $X$.
The averaging is performed over a smaller area within the bulk, away from the surface, in order to capture the specific characteristics of the bulk. 
The details of the averaging process are provided in the Supplementary Materials. 
If the particle fills up homogeneously, then the variance will be zero, whereas any inhomogeneity will increase the value of the integral. 
Therefore, larger the value of the variance, higher is the degree of phase separation. 
\begin{figure*}[t]
\centering
\subfigure[]{\includegraphics[width=2.25in]{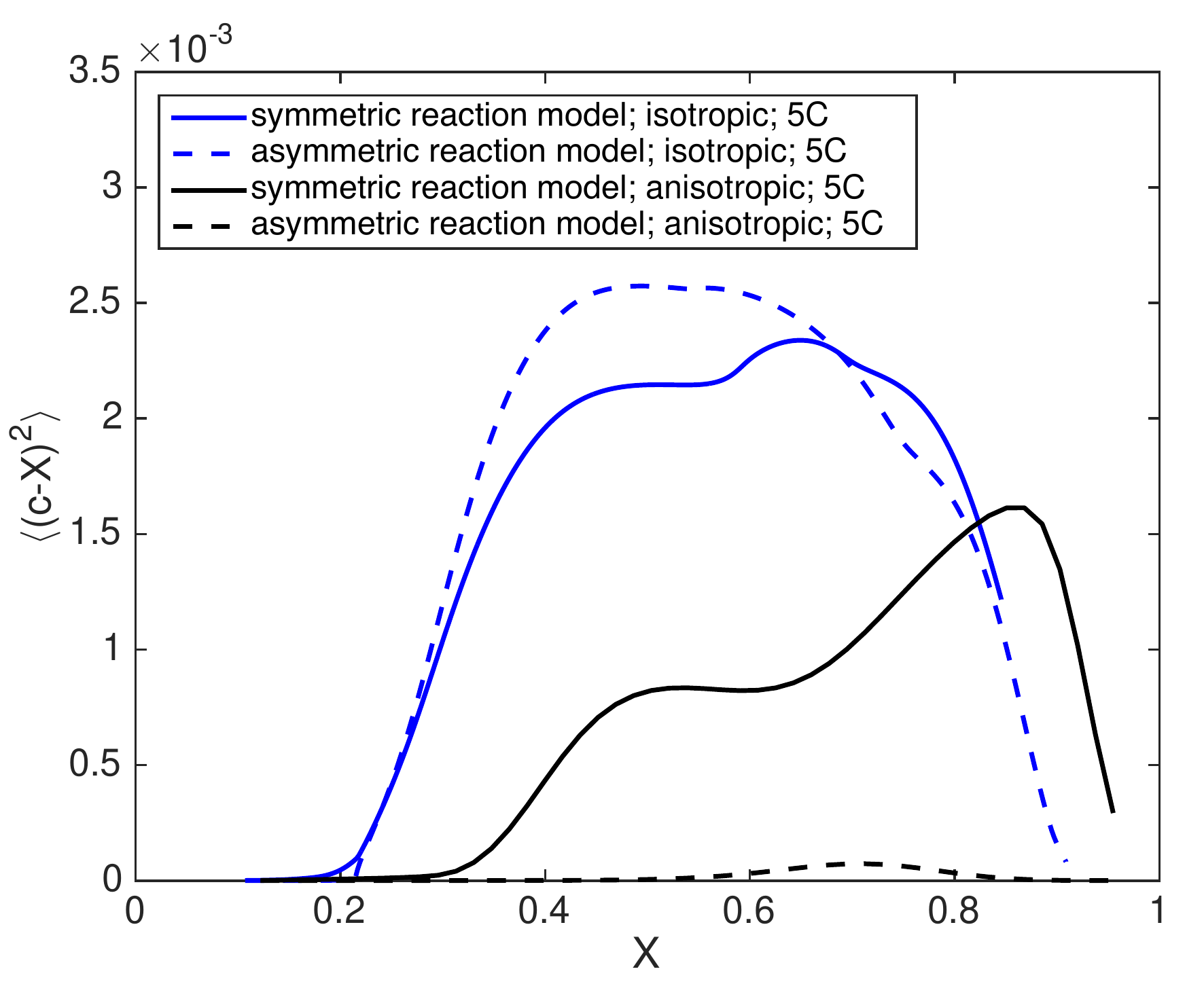}\label{fig:6}}
\subfigure[]{\includegraphics[width=2.23in]{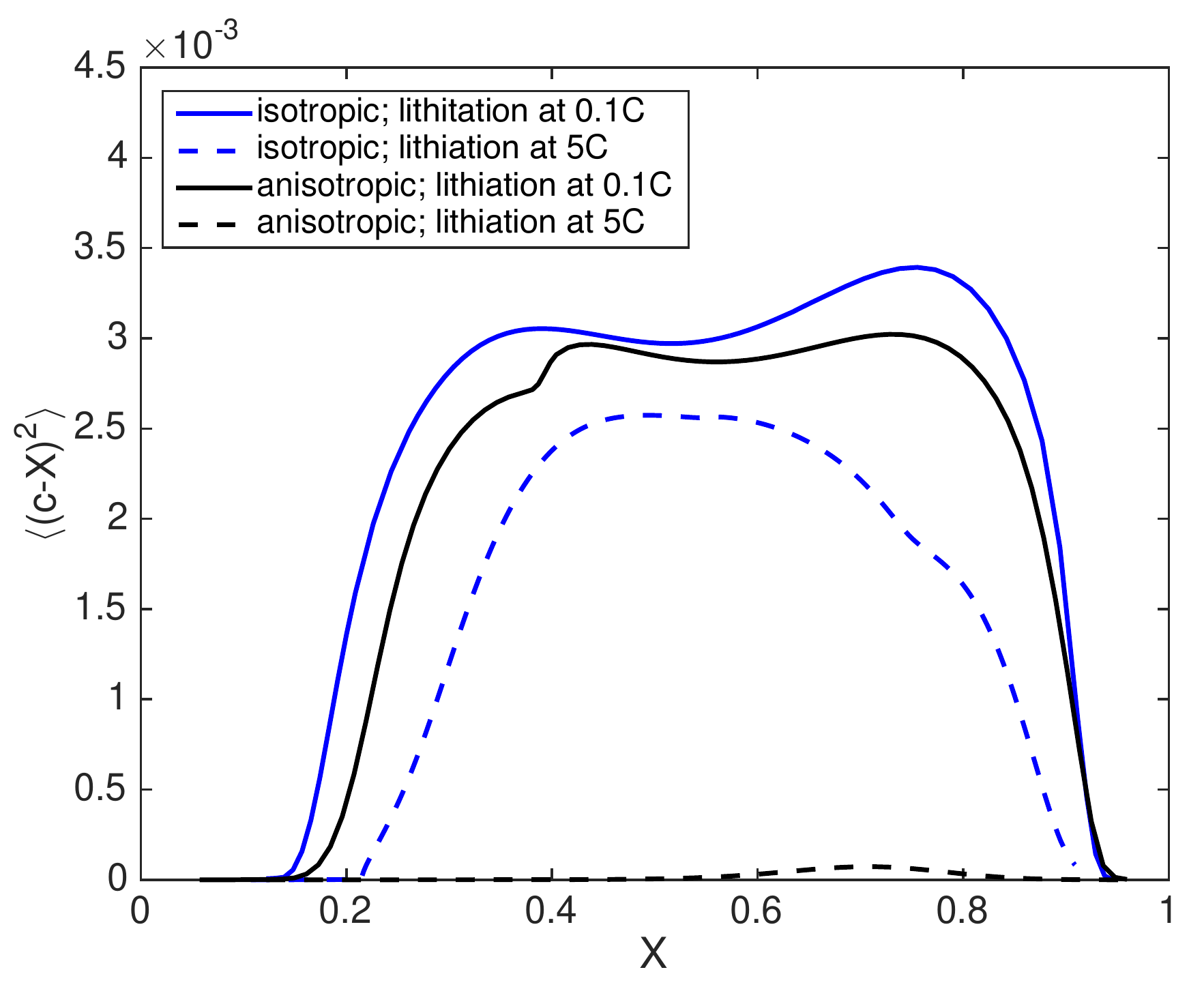}\label{fig:7}}
\subfigure[]{\includegraphics[width=2.3in]{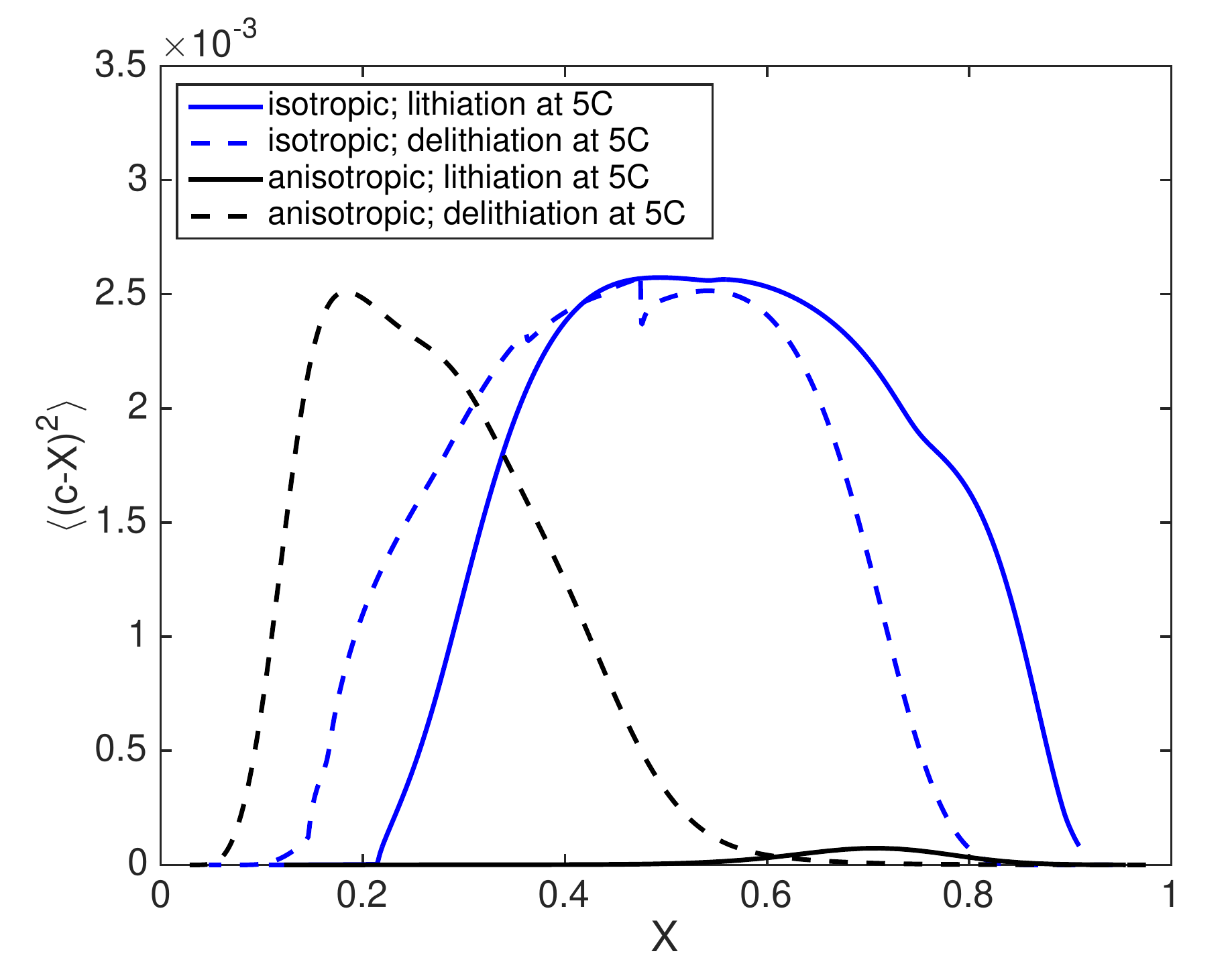}\label{fig:8}}
\caption{Amount of phase separation for the two reaction models: (i) symmetric reaction model~\cite{doyle1993modeling} as given by Eqn.~\eqref{eq:rm_Doyle}, and (ii) asymmetric reaction model~\cite{lim2016origin} as given by Eqn.~\eqref{eq:rm_Lim}. (a) It can be seen that the asymmetry in the exchange current leads to a lower amount of phase separation indicating that phase separation can be suppressed due to the reaction kinetics during lithiation. (b) Lithiation plots for the asymmetric exchange current density for different C-rates and form of interfacial tension tensor. As can be observed, for an anisotropic $\bfkappa$, the asymmetry of the exchange current density has a large effect on the suppression phase separation whereas the suppression is lower for the isotropic case. (c) Lithiation and delithiation plots for the asymmetric exchange current density. As can be observed, for an anisotropic $\bfkappa$, the asymmetry of the exchange current density suppresses phase separation during lithiation and enhances it during delithiation as predicted by the theory~\cite{bazant2017thermodynamic}. In contrast, for an isotropic $\bfkappa$, the phase separation kinetics behavior is relatively unaffected by the reaction model.}
\end{figure*}

To test the theory of driven autocatalytic phase suppression, simulations were performed by varying the reaction kinetics for the same C-rate for the isotropic and anisotropic $\bfkappa$ tensors. 
Fig.~\ref{fig:6} shows the spatial variance for the two reaction models at a much higher C-rate of 5C for an isotropic and an anisotropic interfacial tension tensor. 
For a thin phase boundary in the depth, the scaling analysis predicts that the effect of the surface wetting becomes pronounced,  effectively fixing the concentration at the top and bottom surfaces to be close to the fully dewetted value, causing the exchange current density to be approximately constant.
The almost constant value of the exchange current density has negligible effect on the subsurface phase behavior, thereby producing a similar morphology in both cases.
In contrast, for an anisotropic $\bfkappa$ tensor, the reaction model does affect the phase separation process significantly.
At large C-rates, the phase separation is suppressed by the experimentally determined reaction model corresponding to an asymmetric exchange current density, consistent with theory~\cite{bazant2017thermodynamic}. 
The asymmetry in the exchange current corresponds to an auto-inhibitory reaction during insertion and therefore reduces the degree of phase separation occurring in the bulk~\cite{bai2011suppression,cogswell2012coherency}. 

Next, a comparison of the spatial variances of the concentration profiles for the experimental reaction model with two different C-rates and $\bfkappa$ tensors are shown in Fig.~\ref{fig:7}. 
As the current increases, the lithiation process tends towards the solid-solution regime, which is consistent with previous depth-averaged results~\cite{bai2011suppression,cogswell2012coherency,singh2008intercalation}, and experiments~\cite{lim2016origin}. 
However, the spatial variance is reduced by a significantly larger amount in the anisotropic case. 
The isotropy does not allow for the surface reactions to affect the bulk, hence allowing phase separation even for larger currents as the processes are effectively decoupled.
Experiments show notable suppression of phase separation during lithiation for high C-rates~\cite{lim2016origin}, indicating that the reaction mechanism may be explained by the latter case of an anisotropic $\bfkappa$. 

Finally, in Fig.~\ref{fig:8}, we compare the lithiation and delithiation variances for the experimental reaction model at the large C-rate of 5C, for both, the isotropic and anisotropic cases. 
With isotropy, there is a small difference between the lithiation and delithiation curves. 
The onset of phase separation occurs at different critical concentrations, but the maximum value of the spatial concentration is of the same order during charge and discharge, which implies that the reaction model does not directly affect in the dynamics of phase separation in the bulk. 
However, the plot for the anisotropic interfacial tension tensor shows clear evidence for homogeneous filling and enhanced phase separation during delithiation. 
The reason for this different morphological behavior can be attributed to the fact that, with isotropy, the effects of surface reactions do not penetrate through the bulk thereby giving rise to the same behavior during lithiation and delithiation, whereas, with anisotropy, the bulk behaves like the surface and the mechanism can be explained by the 1D auto-catalytic/auto-inhibitory theory of phase separation, as proposed by Bazant~\cite{bazant2017thermodynamic}.
Therefore, the simulations provide evidence that thermodynamic suppression of phase separation in the depth, during discharge, is necessary for observing the asymmetric phase behavior in the bulk, consistent with experiments in single-crystalline particles. 

To summarize, the anisotropy of the interfacial tension drastically changes the nonlinear dynamics of the phase evolution process in three specific ways: (i) With an anisotropic $\bfkappa$, the reaction model suppresses the phase separation in the bulk in a similar manner predicted by the depth-averaged theory. (ii) A homogenous lithiation pattern is observed for large currents, similar to the behavior seen in depth-averaged models~\cite{bai2011suppression,cogswell2012coherency,singh2008intercalation}, and experiments on single-crystalline particles. (iii) Asymmetric lithiation and delithiation patterns are created by the reaction kinetics at the surface, similar to the ones observed in experiments, i.e. homogeneous filling during lithiation and enhanced phase separation during delithiation.
Another way of analyzing these observations would be to think of the non-local term as introducing a correlation length of the order of the phase boundary thickness~\cite{Cahn1961}. 
The effect of the surface reactions can be felt over a distance equal to the correlation length, and hence reactions impact the bulk when the phase boundary width is of the order of the thickness of the particle. 
This demonstrates that the anisotropy of the interfacial thickness plays a crucial role and may be necessary to explain the non-equilibrium electrochemical lithiation dynamics observed in experiments. 

\section{Validation of the 2D plane-stress model with a full 3D phase field model}

\begin{figure}[h!]
\centering
\subfigure[]{\includegraphics[width=3.2in]{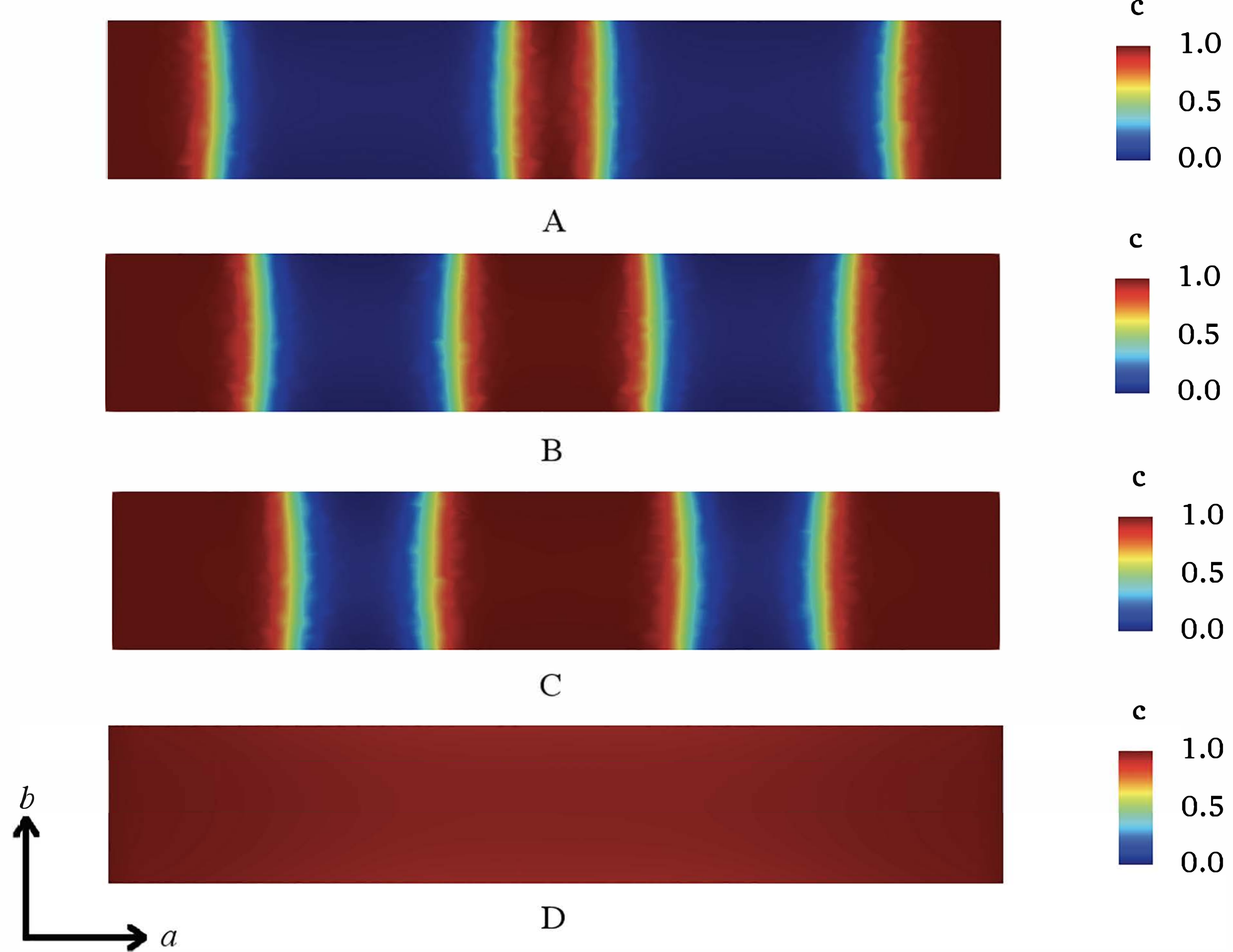}\label{fig:3d_2d_view}}
\subfigure[]{\includegraphics[width=3.2in]{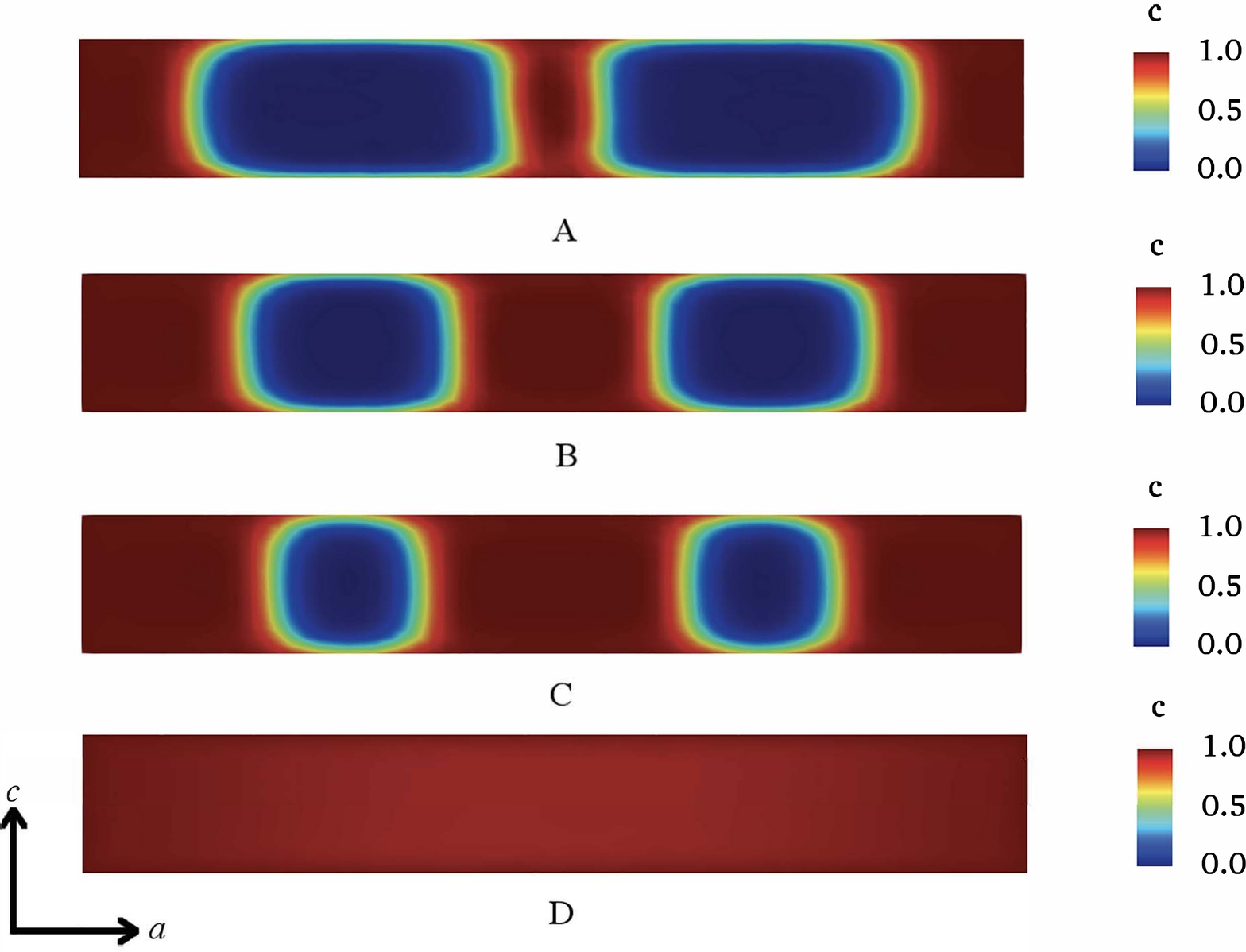}\label{fig:3d_top_view}}
\caption{(a) (A)-(D) Concentration profiles in the $ab$ facet at $z=25\,nm$ during slow lithiation of 1C at SOC's with an anisotropic $\bfkappa$ tensor for an average concentration, $X$ of: (A) 0.3, (B) 0.5, (C) 0.7, and (D) 0.95. (b) (A)-(D) Concentration profiles in the $ac$ facet at $y=15\,nm$ during slow lithiation of 1C at SOC's with an anisotropic $\bfkappa$ tensor for an average concentration, $X$ of: (A) 0.3, (B) 0.5, (C) 0.7, and (D) 0.95.}
\end{figure}

To further validate our findings regarding the anisotropy in the interfacial thickness, along with the validity of the plane-stress assumption, three-dimensional numerical simulations are performed. 
As described earlier, the length of the particle in the $c$-direction was taken to be $25\,n$m and the interfacial energy of the exposed $ab$ planes was considered to be equal to that used for the $ac$ ones. 
The results of the full three dimensional case which take into account different particle shapes and operation conditions will be analyzed in a future study. Figure~\ref{fig:3d_2d_view} depicts the concentration profile for different particle filling fractions.
Additionally, it is known that $D_c = D_a$~\cite{Morgan2004,singh2008intercalation}, leading to a Damk\"{o}hler number of $Da_c = 2\cdot10^{-5}$, a value which indicates the existence of anisotropic diffusion effects in the $ac$-plane as a result of the unequal dimensions of the particle in those directions.

For conciseness, only the case corresponding to 1C is presented. Fig.~\ref{fig:3d_2d_view} depicts the concentration profile for different particle filling fractions. In particular, a sliced view of the $ab$ plane at $z=12.5\,n$m is shown, in order to compare the profile of $\bar{c}$ in the bulk with the 2D case. As has already been discussed in previous sections, and also predicted by the linear stability analysis~\cite{bai2011suppression,bazant2017thermodynamic}, for values of applied current less than the critical one, phase separation cannot be suppressed by the imposed non-equilibrium driving force. When the concentration of lithium in the bulk enters the spinodal region~\cite{bai2011suppression}, a sudden nucleation at approximately $x\sim0.5$ is initiated, resulting in the development of two additional intercalation waves in the bulk~\cite{singh2008intercalation}, Fig.~\ref{fig:3d_2d_view}(A). As time proceeds, it is clear that $\bar{c}$ evolves exactly with the same trend calculated using the plane-stress approximation, fig.~\ref{fig:5}, validating the previous findings. 

The three-dimensional geometric effects are further examined in Fig.~\ref{fig:3d_top_view}, where the $ac$ plane at $y=15\,n$m is presented. The exposed $ac$ facets are affected by the surface wetting properties of the particle, favoring lithium aggregation on these sides. As a result, $\bar{c}$ shows a three-dimensional spatial dependence which is constrained only in a small region with width approximate to $\sim t(\bfe_c)$. Based on equilibrium arguments on phase-separating linear elastic solids~\cite{khachaturyan2013theory}, Cogswell and Bazant~\cite{cogswell2012coherency} showed that the normal of the interface which minimizes the elastic energy of the Li$_X$FePO$_4$ system is $\bfm=[1,0,1]$.  For $X=0.3$, it is found that the left of the newly formed intercalation waves is slightly tilted towards this direction fig.~\ref{fig:3d_top_view}(A). For larger filling fractions though, the orientation of the interface is affected by the imposed non-equilibrium thermodynamic force, decreasing the effects of elasticity on its morphology fig.~\ref{fig:3d_top_view}(B)-(D). This phenomenon ultimately leads to an effective two-dimensional concentration profile. Finally, as shown by Cogswell and Bazant~\cite{cogswell2013theory}, for particles with $50\,n$m width in the $c$-direction, a clear dependence of $\bar{c}$ on both $a$ and $c$ directions was found. Hence, the present study concludes the existence of a critical length under which $\bar{c}$ becomes a weak function of the corresponding spatial direction, leading to the phenomenon of size-dependent miscibility gap~\cite{burch2008phase,Cahn1961} and consequently to suppression of phase-separation.

\section{Conclusions}
In this paper we have presented a thermodynamically consistent electro-chemo-mechanical model for phase separating Li$_X$FePO$_4$ nanoparticles. 
The model accounts for the full coupling between the lithium diffusion process and the elastic deformation and takes into account surface energies which yield wetting/dewetting effects at the nanoparticle surface. 
Non-equilibrium phase-morphologies at different lithiation rates and interfacial tension tensors were studied in the $ab$ plane.
Our 2D phase field simulations, validated by 3D simulations, indicate that the phase boundary anisotropy plays an important role in the subsurface morphology evolution, especially when the phase boundary thickness in the depth becomes of the order of the nanoparticle thickness. 

For the isotropic case, there are two fundamental behaviors observed during lithiation.
The first type, which occurs at low lithiation rates, involves nucleation and subsequent growth of the Li-rich regions near the side facets of the nanoparticle, and is largely consistent with depth averaged models. 
At high rates, in addition to the nucleation and growth from the side facets, a spinodal decomposition occuring at the center of the nanoparticle, creates a phase-interface normal to the $b$-axis, which is not favorable in equilibrium. 
The variation of the concentration in the depth, however, cannot provide an explanation to the experimentally observed phase behavior such as significant suppression of phase separation at high currents and asymmetry in phase behavior during lithation and delithiation.

In contrast, for the anisotropic case, phase separation in the depth is thermodynamically inhibited causing an almost uniform concentration profile across the depth analogous to depth-averaged model predictions, i.e. intercalation waves for small currents and homogeneous bulk solid solution formation for large currents.
As compared to the symmetric exchange current density, the electro-autoinhibitory (asymmetric) nature of the exchange current density significantly suppresses the phase separation for large currents as well as leads to asymmetric lithiation and delithiation phase patterns, in agreement with the theory of driven electro-autocatalysis. 
The simulations strongly indicate the possibility of having an anisotropic phase boundary but further experimentation is necessary to ascertain its exact nature. 


\section*{Acknowledgments}

This work was supported by the Toyota Research
Institute through the D3BATT Center for Data-Driven Design
of Lithium-Ion Batteries.

\bibliography{References}

\end{document}

%% file: defs.tex
\newcommand{\boldface}[1]{\boldsymbol{#1}}  

\newcommand{\bfe}{\boldface{e}}

\newcommand{\bfj}{\boldface{j}}

\newcommand{\bfm}{\boldface{m}}
\newcommand{\bfn}{\boldface{n}}

\newcommand{\bfD}{\boldface{D}}

\newcommand{\bfM}{\boldface{M}}

\newcommand{\bfX}{\boldface{X}}

\newcommand{\bfkappa}{\boldsymbol{\kappa}}

\newcommand{\bfsigma}{\boldsymbol{\sigma}}

\newcommand{\bfnull}{\boldsymbol{0}}
\newlength{\boxwidth}
\def\btheorem{\begin{theorem}}
\def\etheorem{\end{theorem}}
\def\blemma{\begin{lemma}}
\def\elemma{\end{lemma}}
\def\bproposition{\begin{proposition}}
\def\eproposition{\end{proposition}}
\def\bcorollary{\begin{corollary}}
\def\ecorollary{\end{corollary}}
\def\bdefinition{\begin{definition}}
\def\edefinition{\end{definition}}
\def\bexample{\begin{example}}
\def\eexample{\end{example}}
\def\bremark{\begin{remark}}
\def\eremark{\end{remark}}

\newcommand{\be}{\begin{equation}}
\newcommand{\bs}{\begin{split}}
\newcommand{\ee}{\end{equation}}
\newcommand{\es}{\end{split}}
\newcommand{\beq}{\begin{eqnarray}}
\newcommand{\eeq}{\end{eqnarray}}
\newcommand{\bem}{\begin{multline}}
\newcommand{\eem}{\end{multline}}
\newcommand{\ba}{\begin{align}}
\newcommand{\ea}{\end{align}}
\newcommand{\bfig}{\begin{figure}}
\newcommand{\efig}{\end{figure}}